\documentclass[printer]{aa}
\usepackage[varg]{txfonts}
\usepackage{natbib}
\bibpunct{(}{)}{;}{a}{}{,} 
\usepackage{graphicx}
\usepackage{color}

\begin{document}

\title{Signatures of X--ray reverberation in the power spectra of AGN}
\author{I.~Papadakis\inst{1,2} \and T.~Pech\'a\v{c}ek\inst{1,3} \and M.~Dov\v{c}iak\inst{3} \and A.~Epitropakis\inst{1} \and D.~Emmanoulopoulos\inst{4,1} \and V.~Karas\inst{3}}
\institute{Department of Physics and Institute of Theoretical and Computational Physics, University of Crete, 71003 Heraklion, Greece
\and Foundation for Research and Technology - Hellas, IESL, Voutes, GR-7110 Heraklion, Greece
\and Astronomical Institute, Academy of Sciences, Bo\v{c}n\'{\i}~II 1401, CZ-14131~Prague, Czech Republic
\and Physics and Astronomy, University of Southampton, Southampton SO17 1BJ, UK}

\authorrunning{I. Papadakis et al.}
\titlerunning{Reverberation signals in the AGN PSDs}
\date{Received .. ...... 2015/accepted .. ...... 2015}
\abstract{}{To study the effects of X--ray reprocessing in the power spectra (PSDs) of AGN.} {We compute fully relativistic disc response functions in the case of the ``lamp-post" geometry using the full observed reflection spectrum for various X--ray source heights, disc inclination, and spin values of the central black hole. Since the observed PSD is equal to the product of the intrinsic power spectrum with the  ``transfer function" (i.e. the Fourier transform of the disc response function), we are able to predict the observed PSDs in the case of X--ray illumination of the inner disc.}{The observed PSD should show a prominent dip at high frequencies and an oscillatory behaviour, with a decreasing amplitude, at higher frequencies. The reverberation ``echo" features should be more prominent in energy bands  where the reflection component is more pronounced. The frequency of the dip is independent of energy, and it is mainly determined by the black hole mass and the X--ray source height. The amplitude of the dip increases with increasing black hole spin and inclination angle, as long as the height of the ``lamp" is smaller than $\sim 10$ gravitational radii.}{The detection of the  X--ray reverberation signals in the PSDs can provide further evidence for X-ray illumination of the inner disc in AGN. Our results are largely independent of the assumed geometry of the disc-corona system, as long as it does not change with time, and the disc response function is characterized by a sharp rise, a ``plateau", and a decline at longer times.  Irrespective of the geometry, the frequency of the main dip should decrease with increasing ``mean time'' of the response function, and the amplitude of the dip should increase with increasing reflection fraction. }
\maketitle 

\section{Introduction}

It is widely accepted that Active Galactic Nuclei (AGN) are powered by accretion of matter (in the form of a disc) onto a supermassive Black Hole (BH). X--ray emission ($\sim 0.2 - 100$ keV) is ubiquitous in AGN. The main process for their production is generally accepted to be the Comptonisation of thermal disc photons  by the electrons of a hot ``corona" ($kT_{\rm e} \sim 100$ keV). X--rays are thought to be emitted from the innermost parts of the central source, where most of the gravitational power is released. The fast, large amplitude X--ray variations, as well as microlensing observations (e.g. Mosquera et al. 2013), indicate that the X--ray source size is less than 10-15 $r_g$ ($r_g$ is the gravitational radius). Although the exact location and the geometry of the X--ray source are currently unknown, it is possible that part of the X--ray continuum emission is intercepted and reprocessed by matter in the vicinity of the BH. The most prominent features of this reprocessing are the Fe K$\alpha$ fluorescence line at 6.4 keV (for neutral matter) and a broad Compton reflection ``hump'' that peaks at energies $\sim 20- 30$ keV. Intense soft X--ray emission is also expected in the case of mildly ionized discs. 

In the case of X--ray irradiation of the inner disc, general and special relativity effects should leave their mark on the X--ray reflection features. The best studied case is the shape of the emitted iron line which is expected to be asymmetrically broadened (e.g. Fabian et al 1989; Laor 1991; Dov\v{c}iak et al. 2004; Brenneman \& Reynolds 2006; Dauser et al. 2010; Dov\v{c}iak et al. 2014). Furthermore, there should be time delays between the primary X--ray continuum and the reprocessed emission. Soft-band X--ray ``reverberation lags" were indeed discovered with XMM-Newton in the Narrow Line Seyfert 1 (NLS1) galaxy 1H0707-495 (Fabian et al. 2009), and in other bright Seyferts (Emmanoulopoulos et al. 2011; De Marco et al. 2013). High-frequency iron K$\alpha$ reverberation lags have also been detected in a small number of Seyfert galaxies (Zoghbi et al. 2012, 2013; Kara et al. 2013a,b), while similar lags have been reported for the Compton ``hump" in a few sources as well (Zoghbi et al. 2014; Kara et al. 2015).

We present a new method to investigate the disc-corona geometry via the power-spectrum modelling of the X--ray light curves. If X--rays are reprocessed, the signal we observe should be equal to the sum of the X--ray continuum  plus the reprocessed emission. The latter component  is the convolution of the primary emission with the response-function of the disc, for a given geometry (e.g. Cunningham 1975; Karas 2006). In effect, the reprocessed component is a delayed plus ``filtered" version, i.e. a filtered ``echo",  of the continuum signal. In this case,  the X--ray power spectral density functions  (PSDs hereafter) should display features of this ``echo''. These features should depend on the shape and amplitude of the response-function, thus on the physical parameters of the system and the disc-corona geometry (i.e. BH mass, X--ray source size, ``distance" between the X--ray source and the disc, inclination, BH spin). 
 
We use fully relativistic response functions in the case of the ``lamp-post" geometry (e.g. Matt et al. 1991), for various X--ray source heights, inclinations, and spin values of the central BH, to predict theoretically the expected PSDs in the 2--4 and 5--7 keV bands. We discuss the prominent features in the PSDs, and their dependence on the physical parameters of the disc-corona system. The detection of the relativistic ``echo'' features in the X--ray PSDs can provide an additional observational test for the presence of strong relativistic effects in the X--ray emission of AGN. If detected, the parameters of the PSD relativistic features (frequency and amplitude) can be used to determine the parameters of the disc-corona geometry in these objects. 

\section{The disc-response for the lamp-post geometry}

\subsection{The model set up}
We consider the case of a flat disc and a static, point-like X--ray source (the ``lamp") which is located at a fixed point, on the symmetry axis of the system.  We describe below the main assumptions of our modelling (a detailed description of the model can be found in Dov\v{c}iak et al. 2004; 2014).

We assume a Keplerian, optically thick, geometrically thin accretion disc co-rotating around a BH. The disc extends from the radius of the innermost stable circular orbit (ISCO) to 1000 $r_{g}$. We consider three cases for the black hole spin: a non-rotating ($\alpha = 0$),  a fast rotating ($\alpha=0.676$), and a maximally rotating black hole ($\alpha=1$). Since we assume that the disc extends to the ISCO, the spin determines the ISCO radius, hence  the distance between the inner disc and the X--ray source, for a given height. The X--ray source lies above the BH at height $h$. We have considered eighteen source heights, from $h=2.3$ to 100 $r_g$, for all BH spins, plus the height of 1.9 $r_g$ in the case of $\alpha=0.676$, and the height of 1.5 $r_g$in the case of $\alpha=1$.  The disc-corona system is observed by a distant observer, at a viewing angle of $\theta =20, 40$ and 60 degrees (an angle of $\theta=90^{\circ}$ corresponds to a disc seen edge on). In total, we consider 171 different combinations of $\alpha, h,$ and $\theta$. 

The primary emission, $F_{\rm lamp}(E_{\rm intr})$, is isotropic in the local frame, and has a power-law like spectrum with the photon index of $\Gamma=2$, and a high energy cut-off at $E_{\rm cutoff}=300$ keV. We assume that it varies in normalization only, i.e.:
 
\begin{equation}
F_{\rm lamp}(t,E_{\rm intr})=N_{\rm lamp}(t)E_{\rm intr}^{-2}\exp(-E_{\rm intr}/E_{\rm cutoff}).
\end{equation}
\noindent
The observed signal integrated over an energy interval ${\mathcal E}$ is given by the convolution of the variable normalization, $N_{\rm lamp}(t)$, with a linear ``filter/response" function,
\begin{equation}
F_{\rm obs,\mathcal E}(t)=\int\limits_{-\infty}^\infty\psi_{\mathcal E}(t-\xi)N_{\rm lamp}(\xi){\rm d}\xi,\label{Filter_I}
\end{equation}
\noindent where the ``total" response of the system is of the form, 
\begin{equation}
\psi_{\mathcal E}(t)=D_{\mathcal E}\delta(t)+R_{\mathcal E}\Psi_{\mathcal {E},\rm norm}(t).
\end{equation}
The total response function, $\psi_{\mathcal E}(t)$, is equal to the sum of the disc response function, $\Psi_{\mathcal E, norm}(t)$, and the factor $D_{\mathcal E}$, which accounts for the fact that the observed primary emission is in itself  ``filtered". In effect, this factor denotes the total photon count in the energy band $\mathcal{E}$, which reaches the observer.  This is not equal to the emitted photon count in the source's rest-frame, even if the primary source emits isotropically, due to light-bending effects (which are more pronounced when the disc-corona distance decreases). 

The function $\psi_{\mathcal E}(t)$ is a deterministic function which depends on the geometry of the disc-``lamp" system. It does not depend on the ``input" to the ``system" (i.e. on $N_{\rm lamp}(t)$), and it does not vary with time. For physical systems, the observed flux cannot depend on future values of the input, so $\psi_{\mathcal E}(t)=0$ for $t<0$.

\subsection{The disc response-function}
We assume that the X--ray source emits at constant flux (we set $N_{\rm lamp}(t)=1$, in the local frame) for a time period of 
$0\leq t \leq 1t_g$ (time is measured in geometrized units, i.e. $1t_g=4.9255$M$_6$ s, where M$_6$ is the BH mass in units of 10$^6$ M$_{\odot}$). The X--rays illuminate the disc and produce a reflection spectrum. We used the reflection spectrum from a neutral disc computed with the Monte Carlo code NOAR (Dumont et al., 2000) in the constant density slab approximation, assuming solar iron abundance. We assume that the reflection flux is proportional to the incident flux, and that the photons are emitted isotropically in the local frame, co-rotating with the Keplerian disc. 
  
We used a fully relativistic ray-tracing code in vacuum for photon paths from the X--ray source to the observer, from the X--ray source to the disc, and from the disc to the observer. Initially, we compute the observed reflected spectrum from 0.1 up to 8 keV, over energy binns of 20 eV, at time steps of 0.1 $t_{\rm g}$. By summing all the photons that the observer detects (i.e. primary plus reflected component) in an energy interval ${\mathcal E}$ (from $ E_{\rm obs, min}$ to $E_{\rm obs,max})$, at each time $t$, we are able to construct the response-function of the disc, $\Psi_{\mathcal{E}}(t)$,  in this band. We note that we considered the total reflection spectrum that is emitted by the disc and not just the photons that were initially emitted at 6.4 keV (as is often the case with past studies).

\subsection{The disc response in the 5--7 keV band} Figure (1) shows examples of the disc response-functions in the 5--7 keV band (solid lines) for various $h, \theta$ and $\alpha$ values. The $x-$axis measures time in the observer's frame since the beginning of the X--ray source flare (at $t=0$). The response functions are defined in such a way so that $\Psi(t)$ is equal to the ratio of the reflection flux over the normalization of the primary spectrum, $N$.

All response functions share common features: a sharp flux rise at a certain time, $t_{rise}$, a second peak at later times, and a gradual decline after that. The initial rise time corresponds to the time the observer detects first the reflection emission from the near-side of the disc. The second peak appears when the observer detects emission from the far side of the disc. At  longer time scales we detect emission from the outer disc radii, where the reflection amplitude is reduced. Consequently, $\Psi_{\mathcal E}(t)$ decreases to zero.

\begin{figure}
\centering
\includegraphics[width=\hsize]{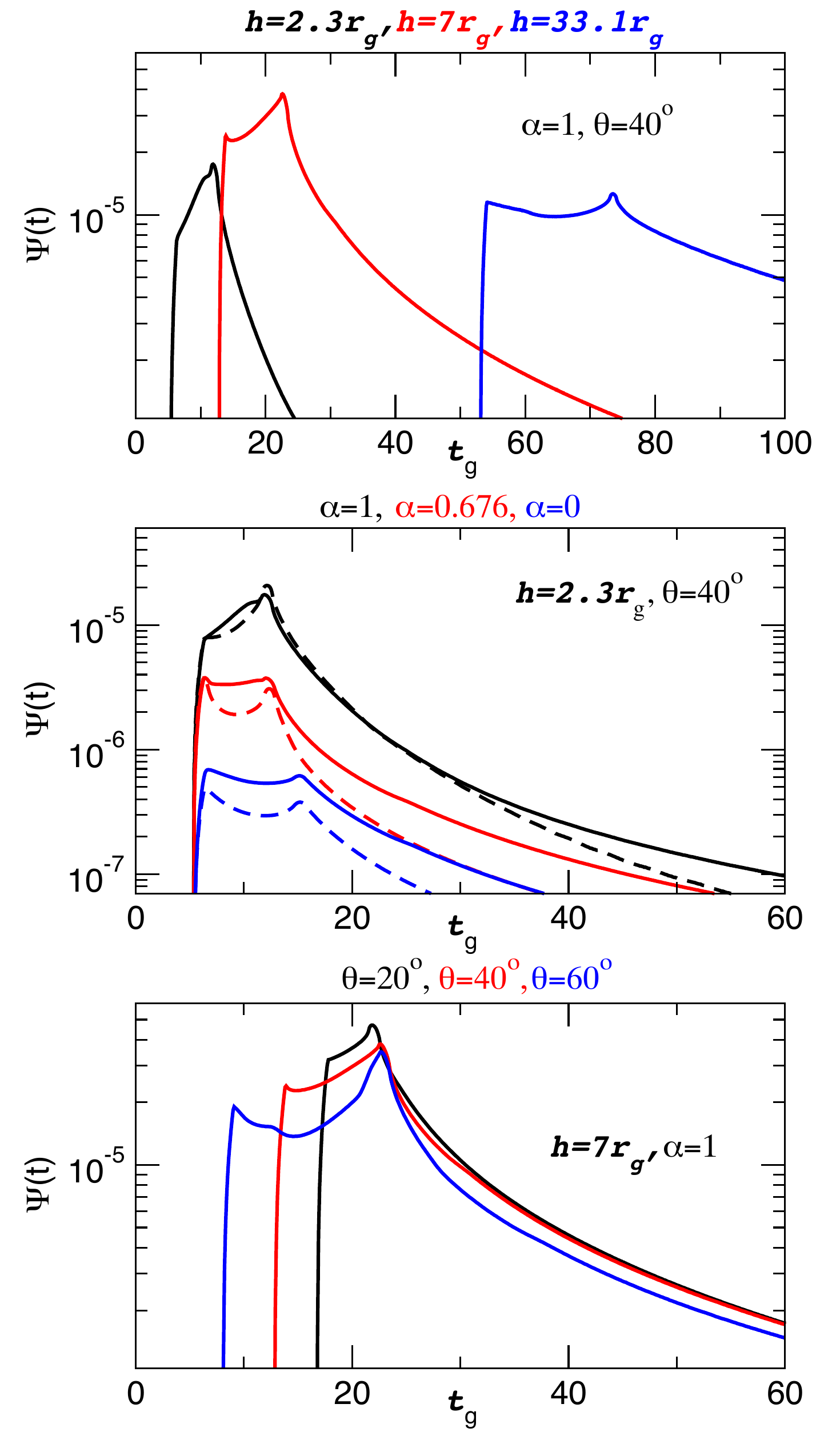}
\caption{Examples of the 5--7 and 2--4 keV band disc response functions (solid and dashed lines, respectively) in the case of the lamp-post geometry, for a $10^7$ M$_{\odot}$ BH and various combinations of the source height, BH spin and inclination angle. Time is measured since the start of the X--ray continuum flash. }
\label{Fig1}
\end{figure}

For a given BH mass, the detailed shape and amplitude of the response function depends on $h,\theta,$ and $\alpha$. For example, the rise time depends mainly on the source's height (it increases with increasing $h$; top panel in Fig.\,1). It also depends on the inclination angle (it decreases with increasing $\theta$; bottom panel in Fig.\,1), but not on the BH spin (middle panel in Fig.\,1). This may seems surprising, but for the height and inclination angles we considered the reflection emission detected first is emitted from a radius larger than $6r_{\rm g}$, irrespective of $\alpha$. The BH spin affects mainly the amplitude of the response function (middle panel in Fig.\,1), specially for small heights. When the X--ray source is located close to the BH, light-bending effects are strong, and most of the X--ray primary emission illuminates the region close to the BH. In the case of a non-rotating BH, the disc does not extend to such small radii and the disc response is significantly reduced. Finally, the width of the response functions is determined mainly by the height and inclination: as $h$ and $\theta$ increase, the time difference between the light path from the far and the near side of the disc increases, and so does the width of the response function.

\subsection{The disc response in the 2--4 keV band} 
The rise time, $t_{rise}$, and the width of the response function do not depend on energy. However, the amplitude of the response function is energy dependent. As an example of this effect, the dashed lines in the middle panel of Fig.\,(1) show the response functions in the 2--4 keV band (for the same $h, \alpha,$ and $\theta$ parameters that we used to compute the 5--7 keV response functions plotted in the same panel). The $\Psi_{2--4 keV}(t)$ amplitude  is always smaller than the $\Psi_{5-7 keV}(t)$ amplitude. 

The response function determines the time average ratio of the reflected component over the total flux (i.e. continuum plus reflection), $f_{refl}$ (reflection fraction, hereafter). The amplitude of the 2--4 keV response function is smaller than the $\Psi_{5-7 keV}(t)$ amplitude because $f_{refl}$(2--4 keV) is systematically smaller  than $f_{refl}$(5--7 keV). This is not only because the reflection component is (intrinsically) weaker in the 2--4 keV band, but also because the primary spectrum increases towards lower energies. 

The middle panel in Fig.\,(1) indicates that the amplitude of the 2--4 keV response function is significantly smaller than the amplitude of the 5--7 keV band response when $\alpha=0$, but the difference reduces with increasing spin. This is because  $f_{refl}$(2--4 keV) increases in the case of maximally rotating BHs and small $h$. In these cases, the X--ray source is located close to the BH and the disc extends to smaller radii. Many iron line photons originally emitted at 6.4 keV are now observed in the 2--4 keV band due to the large gravitational energy shifts. This causes an increase of $f_{refl}$(2--4 keV) and in the amplitude of the disc response in the 2--4 keV band. 

\section{PSD relativistic ``echo''-features in the case of the lamp-post geometry.}

It is well known (see e.g. Section 4.12 in Priestley 1981) that, if the observed signal, $F_{\rm obs, \mathcal E}(t)$, is equal to the convolution of a random function, $N_{\rm lamp}(t)$, with a response function, $\psi_{\mathcal E}(t)$ (i.e. just like in eq.\,2), then its power spectrum, PSD$_{\rm obs, \mathcal E}(\nu)$,  is related with the power spectrum of the ``input", PSD$_{\rm N}(\nu)$, as  
\begin{equation}
{\rm PSD}_{\rm obs, \mathcal E}(\nu)=|\Gamma(\nu)|^2{\rm PSD}_{\rm N}(\nu)/[(D_{\mathcal E}+R_{\mathcal E})^2\langle N_{\rm lamp}\rangle^2],
\end{equation}
\noindent 
where, 
\begin{equation}
\Gamma(\nu)=\int_{-\infty}^{\infty}\psi_{\mathcal E}(t)\exp(-i2\pi\nu t)dt,
\end{equation}
\noindent
is the ``transfer function" of the system. The term $(D_{\mathcal E}+R_{\mathcal E})\langle N_{\rm lamp}\rangle$ in eq.\,(4) is the  mean of the detected flux (as can be easily seen by eqs.\,(2) and 3), and is used so that the observed PSD is normalized to the light curve mean squared.  

Equation (4) shows that the PSD of the ``output" (i.e. of the observed X--rays), at each frequency, $\nu$, depends on the value of the input PSD, PSD$_{\rm N}(\nu)$, and of the transfer function, $\Gamma(\nu)$, at the same frequency. This in contrast to the time-domain relation between ouput-input, e.g. eq.\,(2), where the value of the output at time $t$ depends not only on the value of the input at $t$ but also on past values as well.

\begin{figure}
\centering
\includegraphics[width=\hsize]{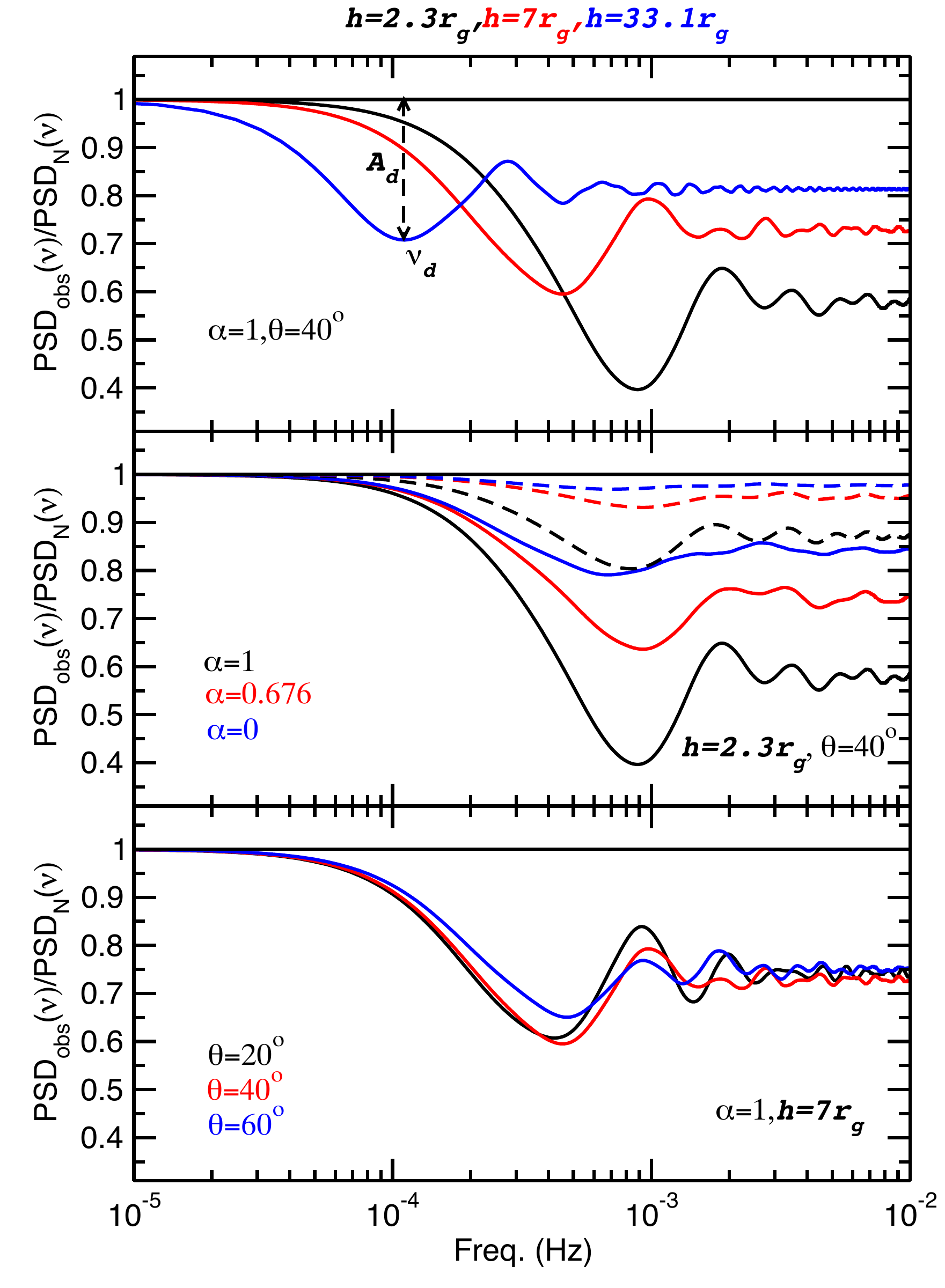}
\caption{The ratio of the observed over the intrinsic PSDs in the 5--7 and 2--4 keV band (solid and dashed lines, respectively), assuming the response functions plotted in Fig\,1.}
\label{Fig3}
\end{figure}

\subsection{PSD signatures of X--ray reverberation in the 5--7 keV band}

Using the disc response functions we presented in the previous section\footnote{Note that the function $\Psi_{\mathcal E,norm}(t)$, which appears in the second part of eq.\,(3), is a normalized version of the disc response functions we discussed in \S2.2 and 2.3, in such a way so that its integral over all times is equal to unity. Its amplitude is determined by the constant $R_{\mathcal E}$.}, and the factor $D_{\mathcal{E}}$, which we computed for all heights, inclination angles, and BH spin values we consider, we can estimate the total response function from eq.\,(3), and then the transfer function from eq.\,(5). The solid lines in Fig.\,(2) show the ratio of the observed over the intrinsic PSDs in the 5--7 keV band, for the same model parameters that we considered in Fig.\,(1). This ratio is independent of the form of the intrinsic PSD, as it is always equal to $|\Gamma(\nu)|^2$ for a fixed disc-corona geometry. The PSD ratios plotted in this figure are estimated for a BH mass of 10$^7$ M$_{\odot}$, using eq.\,(4). 

The PSD ratios are smaller than unity at high frequencies. This indicates a loss of power in the observed PSD at high frequencies, as expected. Interestingly, this ``power loss" has a distinctive shape: the main feature in the PSD ratios is a prominent dip at frequencies higher than $\sim 10^{-4}$ Hz. At requencies higher than $\sim 1-2\times 10^{-3}$ Hz, the ratios show an oscillatory behaviour around a value which is equal to $[D_{\mathcal E}/(D_{\mathcal E}+R_{\mathcal E})]^2$ (in the case when the PSDs are normalized to the mean squared). 

\begin{figure}
\centering
\includegraphics[width=\hsize]{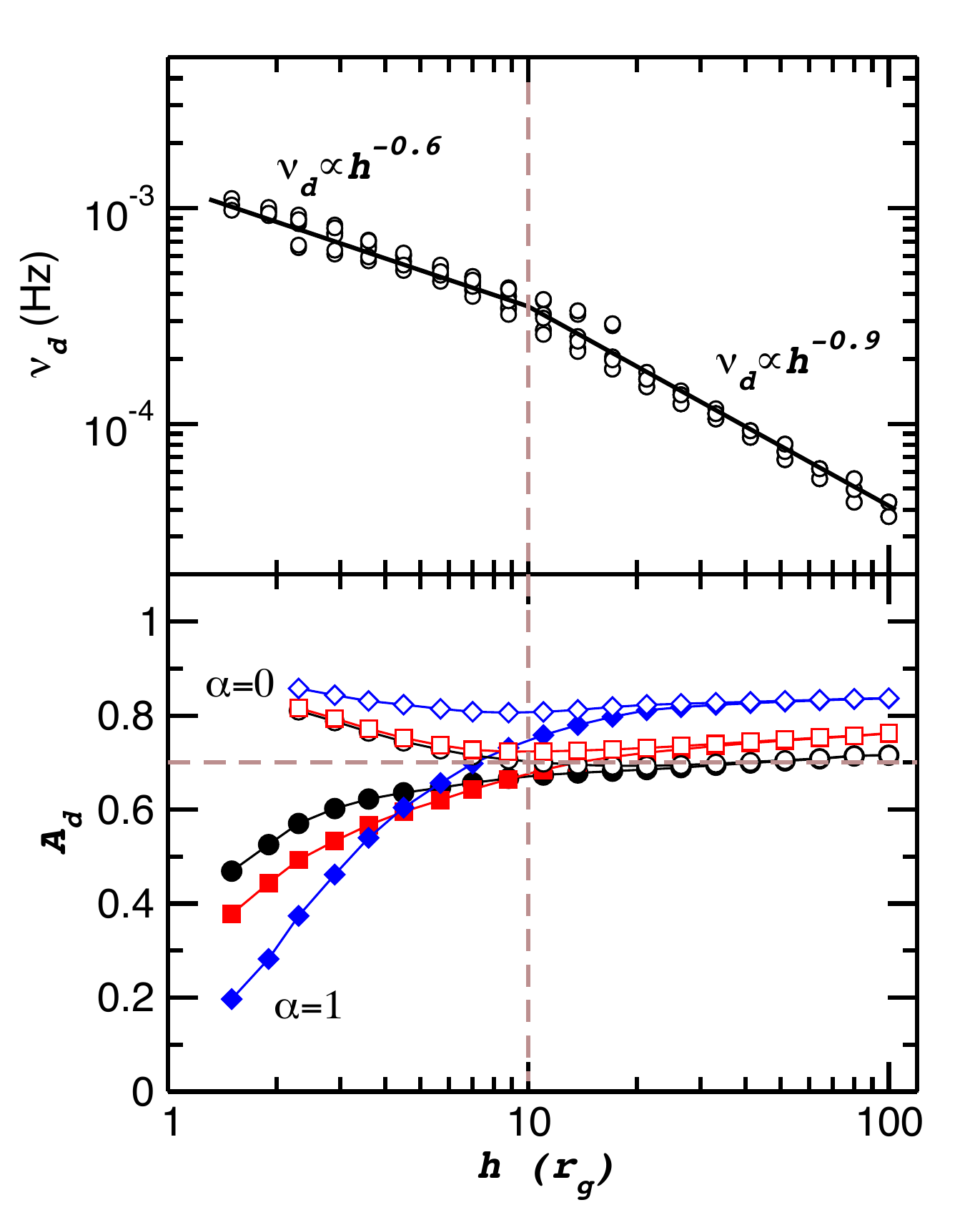}
\caption{The frequency, $\nu_d$, and the amplitude, $A_d$, of the maximum depression of the first dip in the PSD ratios, plotted as function of the source height (top and bottom panel, respectively). Open and filled points in the bottom panel indicate the data for the $\alpha=0$ and $\alpha=1$ case, respectively. Black, red and blue points indicate the data for $\theta=20, 40,$ and $60$ degrees, respectively.}
\label{Fig4}
\end{figure}

\begin{figure}
\centering
\includegraphics[width=\hsize]{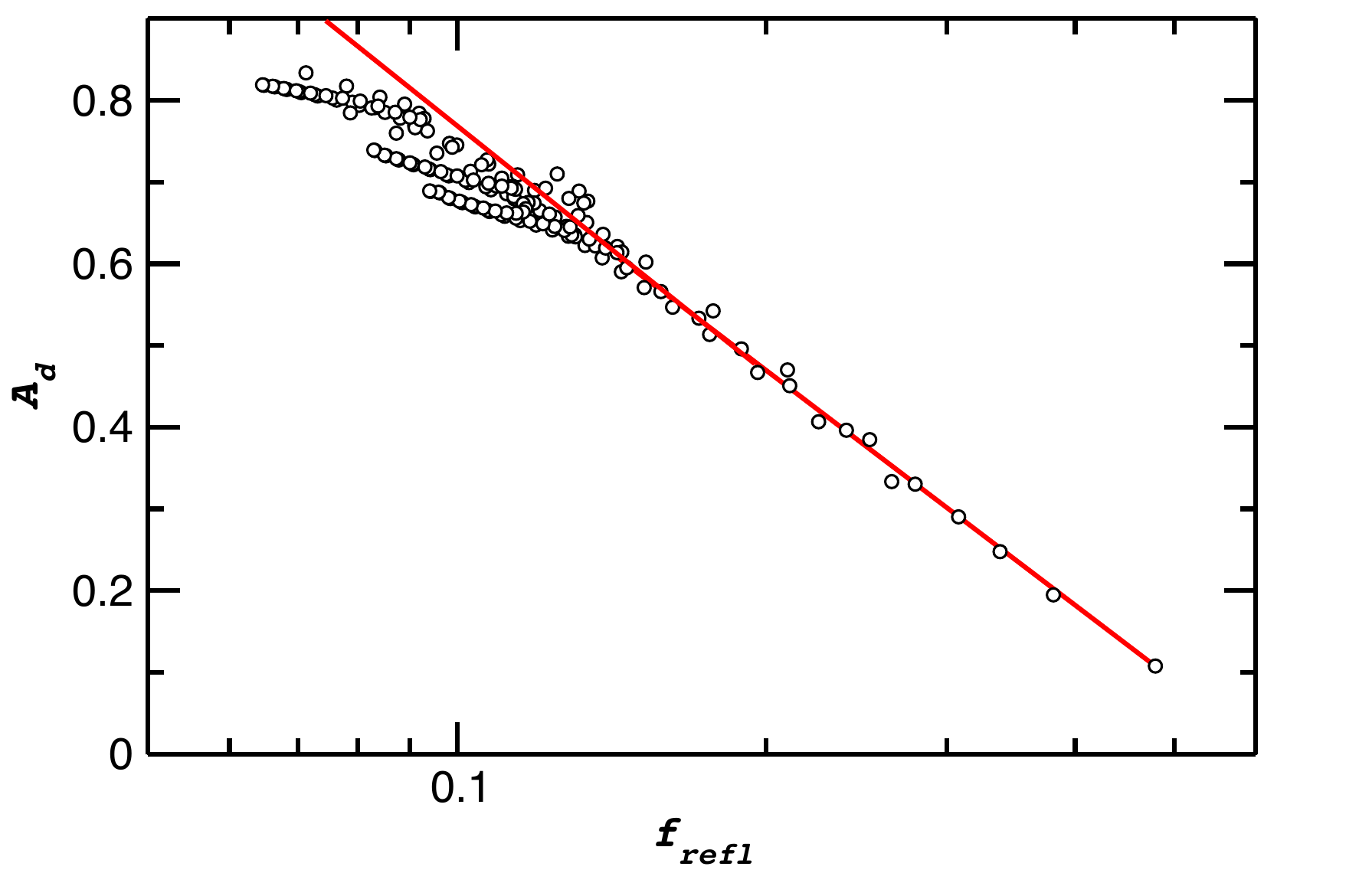}
\caption{The amplitude, $A_d$(5--7 keV), of the maximum depression in the first dip of the PSD ratios plotted as function of $f_{refl}$(5--7 keV).}
\label{Fig5}
\end{figure}

The top panel in Fig.\,(3) shows the frequency of the maximum depression in the first dip, $\nu_d$, versus $h$, for all the parameter combinations we considered in this work. This plot shows that $\nu_{d}$ depends strongly on the source height: as the height increases, $\nu_d$ decreases. The solid lines indicate the relations $\nu_d=1.4\times10^{-3}h^{-0.58}$ and $\nu_d=3\times10^{-3}h^{-0.93}$, which are plotted below and above $h=10 r_g$, respectively. There is no significant scatter of the data around these relations. This result indicates that $\alpha$ and $\theta$ do not affect $\nu_d$ significantly. If such a feature is detected in the observed PSD of a source, then the source height can be estimated from $\nu_d$.

The amplitude  of the main dip, $A_d$, depends on the BH spin when $h$ is smaller than $\sim 10 r_g$. For example, when $h=2.3 r_g$ (middle panel in Fig.\,2), the dip amplitude increases significantly with increasing $\alpha$, reaching values as low as $\sim 0.4$ in the case of maximally rotating BHs. The amplitude depends on the inclination angle as well. In the bottom panel in Fig.\,(3) we plot $A_d$(5--7 keV) as a function of $h$ in the case of $\alpha=1$ and $\alpha=0$ (filled and open points, respectively), for all inclinations we considered (circles, squares and triangles indicate the $\theta=20^{\rm o}, 40^{\rm o},$ and $60^{\rm o}$ data). At heights larger than $\sim 10 r_g$, $A_d$ is more or less independent of the BH spin, and decreases with increasing inclination angle. However, at lower heights, $A_d$ depends strongly on $\alpha$. If $\theta$ is known, and $h$ is determined from the $\nu_d-h$ relation, then the spin value can be estimated from $A_d$, specially at low heights. In this case, if $A_d$ is observed to be smaller than $\sim 0.7$ (indicated by the horizontal dashed line in the bottom panel of Fig.\,3), this would indicate a rotating BH.

In general, the amplitude $A_d$ depends strongly on $f_{refl}$. This is shown in Fig.\,(4), where we plot $A_d$(5--7 keV) versus $f_{refl}$(5--7 keV). Clearly, $A_d$ decreases (i.e. the dip becomes deeper) as the reflection component in the 5--7 keV band increases. The solid line in this figure indicates the relation $A_d=-0.2-0.4\ln (f_{refl}$), which describes well the overall relation between $A_d$ and $f_{refl}$ when $f_{refl} \gtrsim 0.13$. At lower reflection fractions (i.e. large heights, small inclination angles, and low BH spins) $A_d$ still depends on $f_{ref}$, but the relation between $A_d$ and $f_{refl}$ becomes less steep.

\subsection{PSD signatures of X--ray reverberation in the 2--4 keV band}

A similar loss of variability  at high frequencies  is also expected to be detected in the observed 2--4 keV PSDs. As above, we expect to detect a dip in the ``observed over the intrinsic" 2--4 band PSD ratios, and then an oscillatory  behaviour, with a decreasing amplitude at higher frequencies. As an example, the dashed lines in the middle panel of Fig.\,(2) show the expected 2--4 keV band PSD ratios in the case of $h=2.3r_g$, $\theta=40^{\rm o}$ and the three spin parameters listed in that panel. The frequency of maximum depression for the first dip is almost identical in the 2--4 and 5--7 keV PSD ratios. This is true in all cases. On the other hand, $A_{d}$ is much more pronounced in the 5--7 keV, rather than the 2--4 keV ratios. This is explained by the fact that the reflection component in the 5--7 keV band is much stronger than in the 2--4 keV, as we explained in \S2.3. In fact, for spin parameters smaller than 1, the 2--4 keV PSD ratios show a relatively low amplitude suppression at high frequencies, which is almost equal to zero for the case of a non-rotating BH (middle panel of Fig.\, 2). 

\section{Discussion}

We presented the results from the study of X--ray reverberation effects in the power spectra of AGN. In general, if X--ray reflection of the inner disc operates in AGN, and  if the X--ray primary continuum is variable, the X-ray reflection component will be a delayed plus``filtered" version of the continuum signal. The X--ray reflection component should have a variability amplitude smaller than the primary emission, and the power spectra of the observed light curves, which are the sum of the direct and the reflection components, should show a loss of variability power at high frequencies.

\subsection{Summary of our results}

Our results are based on the hypothesis of the lamp-post geometry,  mainly because it is relatively straight forward in this case to predict theoretically the disc response to X--ray reflection. This geometry  has been adopted by many authors in the past when studying the variability properties of the X--ray reflection component (e.g. Reynolds et al. 1999; Miniutti \& Fabian 2004;  Nied\v{z}wiecki \& Miyakawa 2010; Emmanoulopoulos et al. 2014; Cackett et al. 2014). From a physical point of view, the lamp-post geometry can account for the main features of the X--ray source if the latter  is the site where an outflowing corona is accelerated for example, or where the shocks in an aborted jet collide (Ghisellini et al. 2004).

We found that the expected loss of variability power at high frequencies follows a distinctive and characteristic pattern: the observed PSDs should show a prominent dip and an oscillatory behaviour, with decreasing amplitude, at higher frequencies. The frequency of the dip is energy independent, but the amplitude of the prominent dip in the observed PSD, $A_d$, increases with increasing reflection fraction. Therefore, the X--ray reverberation PSD ``echo" features should be more prominent in energy bands where the reflection component is strong. Indeed, according to our results, the reverberation PSD features are more prominent in the 5--7 keV than the 2--4 keV bands.

In the case of the lamp-post geometry, the central frequency of the prominent dip, $\nu_d$,  depends mainly on the X--ray source's height, $h$,  for a given BH mass\footnote{We considered a BH with a mass of $10^7$ M$_{\odot}$ in our study. Our results are valid for any BH mass, provided that $\nu_d$ is multiplied by a factor of $10^7/$M$_{\rm BH}$, where M$_{\rm BH}$ is the mass of the BH in solar units.}. It decreases with increasing $h$, as $\nu_d\propto 1/h$ or $\nu_d\propto 1/\sqrt{h}$, in the case of $h>10$ and $h<10 r_g$, respectively. If we can determine $\nu_d$, we can therefore infer the X--ray source height. The amplitude of the dip depends on the BH spin and the inclination angle.  If $h$ is smaller than $10 r_g$,  the BH spin and the inclination angle can be determined from $A_d$. 

\subsection{Implications of alternative X-ray/disc geometries}

 The hypothesis of a ``point-like" source located at (a fixed) height $h$, on the disc symmetry axis is an over simplification of the geometry of the X--ray source, which may be complex, non-spherical, and of finite dimensions.  In addition, the disc may be warped, or its thickness may be increasing with the radius. However, our main results should not depend strongly on the exact disc-corona geometry. In almost all the above geometry configurations, the disc response function should increase sharply at $t_{rise}$, it should show a ``plateau", and then it should decrease to zero at longer time scales. For example, the disc response function in the case of extended X--ray sources do show these features (Wilkins \& Fabian, 2013). The transfer function in this case (and hence the observed PSD as well) will show a prominent dip, characterized by its central frequency, $\nu_d$, and amplitude, $A_d$, and an oscillatory behaviour,  with decreasing amplitude, at high frequencies, just like in the lamp-post geometry. 

\begin{figure}
\centering
\includegraphics[width=\hsize]{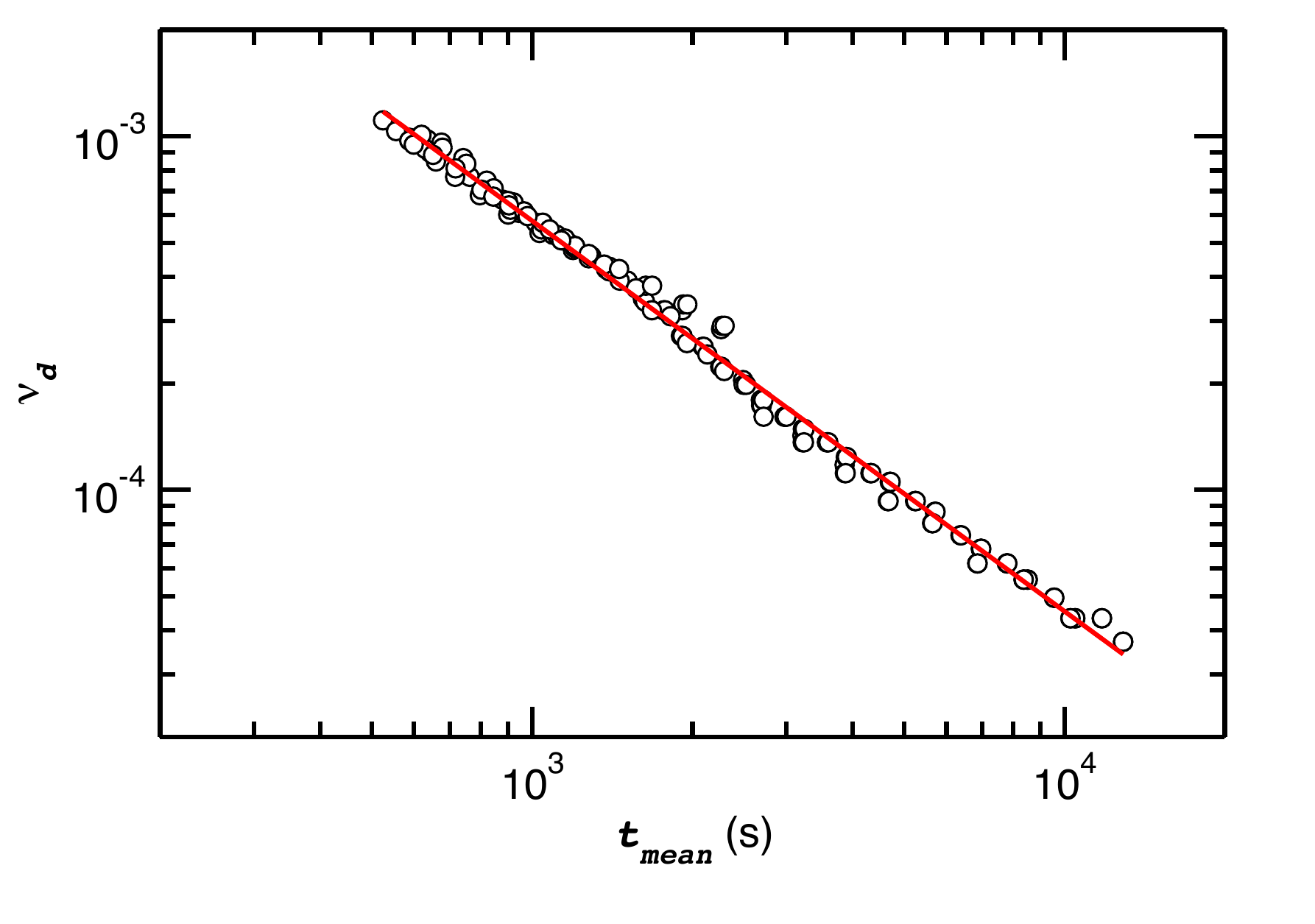}
\caption{The frequency $\nu_d$ plotted as a function of the ``mean time of response", $t_{\rm mean}$.}
\label{Fig5}
\end{figure}

We believe that the dependence of the dip's amplitude on $f_{refl}$ (Fig.\,5) should still hold under any geometry, because the amplitude of the response function should increase(decrease) with increasing(decreasing) reflection fraction. However, the $\nu_d-h$ relation (top panel in Fig\,4) will not hold in a geometry different from the lamp-post geometry. To investigate this issue further, we estimated the ``mean time of response" as: $t_{\rm mean,\mathcal E}=\int\limits_0^\infty t \Psi_{\mathcal E}(t)  {\rm d}t$, for the case of the lamp-post response functions we discussed in \S 2.2. This time scale depends on $t_{rise}$, as well as on the width of $\Psi_{\mathcal E}(t)$. 

Figure (6) shows a plot of $\nu_d$ as a function of $t_{\rm mean}$ in the 5--7 keV band,  for all the lamp-post model parameters we considered. The correlation is almost perfect. The solid line indicates the relation: $\nu_d=1.2\times t_{\rm mean}^{-1.1}$, which fits the data well. In comparison to the similar plot shown in the top panel of Fig.\,(4), the scatter of the data points around the model line is significantly reduceded, and we do not detect a ``break" in the $\nu_d-t_{\rm mean}$ relation, as we did in the $\nu_d-h$ relation. This is most probably due to the fact that $t_{\rm mean}$ depends both on $\theta$ and $h$, hence the simple $\nu_d\propto t_{\rm mean}^{-1}$ relation we ended up with. We suspect that the same relation will hold for any response function with the features we mentioned above. 

The above are valid as long as the X--ray source/disc geometry does not vary with time. 
For example, an outflowing/inflowing X--ray corona atop the accretion disc will result in a time-varying disc reponse. This should introduce PSD features which will correspond to the time scales characterizing the evolution of the dynamic corona. An additional geometrical configuration which can lead to a variable disc response and transfer function is the one where the X--ray reflection pattern is non-homogeneous.  This situation could arise in the case when the disc is illuminated by a bright, localized ``active region" at a particular radius, which appears and disappears on a certain time-scale. In this case, X--ray reverberation can even increase the observed PSD amplitude at high frequencies (Pech\'a\v{c}ek et al, in prep). 

We conclude that, irrespective of the details of the exact disc-corona geometry, as long as this geometry does not vary with time, X--ray reverberation should cause a prominent dip to appear in the observed PSDs at high frequencies, and an oscillatory behaviour at even higher frequencies. The detection of these reverberation ``echo" effects in the observed PSDs, at energies where the reflection component contributes significantly in the observed flux, can provide a direct confirmation of the X--ray illumination of the inner disc in AGN. In addition, accurate PSD modelling can provide estimates of the disc-corona system, in a way similar to the  modelling of the iron line shape and of the time-lags spectra. Under the hypothesis of the lamp-post geometry, $\nu_d$ and $A_d$ can be used to determine the source's height, and the BH spin and inclination angle (in the case when the source's height is smaller than $\sim 10 r_g$), using the results we presented in this work. In general, the central frequency and amplitude of the prominent dip in the PSDFs can be used to estimate the ``mean time of response" (via $n_d$) and the time average reflection fraction (via $A_d$), irrespective of the assumed source/disc geometry. 
 
\subsection{The disc ionization and the iron abundance}

Our results are applicably to 2--4 and the 5--7 keV band PSDs only, in the case of X--ray reflection from a neutral disc. We plan to compute the disc response functions in the case when the accretion disc is ionized, however, we expect the same feature (i.e. a prominent high frequency dip and an oscillatory behaviour at higher frequencies) to appear in the soft band PSDs as well. As we demonstrated in \S3.2, the frequency of the dip is energy dependent (as it depends on the geometry only). However, the dip amplitude should be different, as it depends on the strength of the reflection component (see Fig.\,4). In fact, if the reflection component is significant in the soft band, we expect an even stronger dip, which will make it easier to search for the X--ray reverberation``echo" signals. Our results depend on the assumption of solar iron abundance. An iron abundance higher than solar will mainly affect the amplitude of the response function, hence, the amplitude $A_d$ of the prominent dip in the PSDs. In fact, a larger iron abundance will probably result in a ``deeper" $A_d$, which should be easier to be detected. The opposite should be true for a sub-solar iron abundance. 

\subsection{Detection of the relativistic PSD "echo" features in practice}

We plan to study the X--ray PSDs of many X--ray bright AGN (using archival data) to search for the X--ray reverberation ``echo" features. We will present the results from this study in a  forthcoming publication. The best targets should be sources where strong and asymmetric iron lines, as well as reverberation time-lags, have been reported in the literature. Many of the potential targets are of known BH mass (from optical reverberation studies), so we will be able to search for the prominent ``echo" dips at frequencies known in advance. Our results show the PSD ``echo" features in the ratio of the observed over the intrinsic PSD, which is not known a priori. The observed PSDs in the 2--10 keV band generally follow a ``bending power-law" like shape, with slopes roughly equal to unity below a bend-frequency, and $\sim 2$ at higher frequencies (e.g. Gonz\'{a}lez-Mart\'{i}n \& Vaughan 2012, and references therein).  One could perhaps fit the 5--7 keV band PSDs with similar models, and investigate the best-fit residuals as carefully as possible to search for the ``echo" features. On the other hand, as we showed, similar features are also expected in the 2--4 keV PSDs, but with significantly smaller amplitude. Given the proximity in energy of the 2--4 and 5--7 keV bands, the respective intrinsic PSDs may not be drastically different. Perhaps one could fit the 2--4 keV band PSDs, and use these best-fits as proxy for the intrinsic 5--7 keV band PSDs. Perhaps the best-way to search for the PSD relativistic ``echo" features is to study the PSDs at energies below 1 keV, where the reflection fraction may be large, and the signal-to-noise ratio of the observed light curves is much larger than in the 5--7 keV band (for data provided by current satellites, like {\it XMM-Newton} for example).

\begin{acknowledgements} 
This work was supported  by the "AGNQUEST" project, which is
implemented under the "Aristeia II" Action of the "Education and Lifelong Learning"  operational programme of the GSRT, Greece. 
\end{acknowledgements}

\bibliographystyle{aa} 
\bibliography{os} 

\end{document}